# A Conceptual Framework to Analyze Enterprise Business Solutions from a Software Architecture Perspective

**Basem Y. Alkazemi**
Collage of Computer and Information Systems
Umm Al-Qura University
Makkah
Saudi Arabia
*bykazemi@uqu.edu.sa*

**Abstract**
The architectural aspects of software systems are not always explicitly exposed to customers when a product is presented to them by software vendors. Therefore, customers might be put at a major risk if new emerging business needs come to light that require modification of some of the core business processes within their organizations. So they might need to replace their existing systems or re-architect old ones to comply with new architectural standards. This paper describes a proposed framework that helps organizations to build a comprehensive view of their system architecture prior to dealing with vendors. Consequently, every organization can have a reference model that facilitates negotiation and communication with software vendors. The paper applies the proposed framework to an organization in the region of Saudi Arabia to validate its applicability and generates an architectural design for their software systems.

## 1. Introduction

Many software vendors describe their products from a business perspective in a manner to sell only. The only things that are described to customers are the functional aspects of the systems. However, architectural details that express how their system is structured are not explicitly defined by vendors. One possible reason for this is the lack of knowledge regarding software architecture's significant impact upon business needs. In fact, it is very rare to find an organization that has a plan for adopting an extensible architecture for their software systems prior to looking for products in the market. They usually look for vendors that satisfy their business needs within the timeframe and budget available to them, with no regard for how these systems are going to be built and what the potential consequences of adopting a specific vendor's technology might be.

Our observation to a number of organizations across the region of Saudi Arabia concluded that many of them employ different technologies in their systems to satisfy a number of common business needs. For example, some may use Oracle E-Business Suite for their employment management systems while they use Microsoft SharePoint for their website. Others may use PHP for their website and SharePoint for their intranet applications in addition to Oracle forms for financial and warehousing applications. Although the variety of technology within an organization is usually unfavorable as far as management is concerned, this variety might be beneficial to increase flexibility and extensibility of the business needs for an organization. However, it would not be feasible to apply this advantage in practice unless the organization has a solid architecture that describes different layers where every aspect of functionality may fit.

This paper is designed to draw organizations' attention within the region of Saudi Arabia towards the importance of planning for their IT projects from an architectural perspective in addition to the business needs as that seems to be the part that is lacking in many IT projects in the region. It argues that understanding architectural specifications in addition to the functional ones is important, especially in cases where organizations need to ensure flexibility, extensibility and consistency of their systems. Therefore the components of their systems that may need to be modified, extended, or replaced can be identified and managed more practically. This paper describes a proposed architecture for an enterprise system and uses this architecture as a framework to evaluate some common enterprise solutions in the Saudi Market. We selected Enterprise Resource Planning (ERP) [4] as a system to evaluate against our framework from an architectural perspective. One reason for selecting such a system is that ERP is commonly known as a software system that manages the different business applications within organizations. There is no survey in the existing literature that discusses the dimensions we described in this paper as the base of comparison between different ERP vendors. Most of the surveys are based on attributes such as functional capabilities, usability, cost, technology used and customer satisfaction rather than architectural features. Moreover, this paper establishes the basis for achieving comprehensive alignment between business improvement and software architecture activities that is always lacking among enterprises [17].

The remainder of this paper is organized as follows. Section 2 presents a background discussion about software architecture to set up the context of the work. Section 3 describes the key quality attributes from a systems perspective. Section 4 presents the proposed architectural layer of an enterprise solution. Section 5 discusses the main features of a number of ERP solutions in the market. A case study that reports the utilization of our framework to generate a system architecture for an organization is described in section 6. Section 7 presents migration roadmap of UQU systems to comply with our framework. Finally, the conclusion and possible recommendations are given in section 8.

## 2. Background Review on Software Architecture

People usually refer to the term 'architecture' to indicate the physical construction of a building in terms of external shape, and also how the rooms are structured within that building. In software, the word 'architecture' is a term that is in general use, with a number of different interpretations. However, as an analogy to its meaning in civil engineering, it inspires the meaning of creating a product (a software system in this case) from a number of selected components rather than building a single monolithic one. So the way components must be incorporated, the orders in which they must be placed, and the mechanism of interaction between them, are parts of what a system architecture describes.

Bas et al. [7] defined software architecture as the structure of a system that comprises software elements, their external visible characteristics, and the relationship between them. IEEE 1471 [8] defines software architecture as "the fundamental organization of a system embodied in its components, their relationship to each other's and the environment, and the principles guiding its design and evolution". Jones [9] defined architecture as the structure that is composed of components and rules that establish the basis for the interaction between them. All the definitions agree that architecture is concerned with the constituting parts of a system and the relationship between them.

In the literature, many of the available works have explained the significance of considering architecture in software systems. One reason for considering software architecture is to help our understanding of complex software systems. Shaw and Garlan [10] suggested that architecture can be used to define the overall design of a system. Garlan and Perry [11] identified the benefits of considering software architecture in software development as providing support for re-using, evolving, analyzing, and managing software. Budgen [12] considered software architecture to be a way of describing the constructional aspects of a software system at a high level of abstraction (e.g. design stage). Allen [13] identified architecture as being the vehicle to communicate between the requirement and the implementation stages. Szyperski et al. [14] suggested that architecture is important for establishing a context for software systems representing standards and platform requirements.

Garlan et al. [15] identified a number of architectural characteristics that might cause a mismatch to occur in terms of component interaction within a system. These characteristics are:

- The infrastructure that a component is primarily built on.
- Control issues of whether a component can generate a control signal or not.
- The data type manipulated by a system and the way it is transferred between components.
- The pattern of interaction between components.
- The sequence that components must be instantiated and invoked with.

Yakimovitch et al. [16] refined the work of Garlan and identified five variables that describe assumptions about components' interactions, namely packaging, control, information flow, synchronization, and binding. Their main motivation was to establish a mapping between architectural assumptions and a number of problem domains that conform to certain standard architectural types. They demonstrated that the defined variables can be used to abstractly classify different software architectures.

All of the above-presented work emphasizes the importance of considering software architecture as a vehicle to fully understand the different parts of a system. This can help organizations to fulfill their business needs. In fact, considering software architecture is significant to organizations as it helps them to identify whether or not a functional component can be seamlessly integrated into their system without interrupting their daily working routine. In addition, the system must be able to accommodate possible growth in an organization's business. As a result, a number of attributes must be satisfied by software systems to ensure the readiness of such a system as the business grows. The next section discusses a number of key quality attributes that establish the context for evaluating a vendor's solutions.

## 3. System Quality Attributes

In the context of software engineering best practices, an enterprise software system must satisfy a number of key quality attributes that will ensure its readiness to accommodate new business needs without affecting its overall software architecture. So, a system adhering to these attributes can be considered a healthy system to accommodate emerging business needs. These attributes include:

- *Reference schema*: tables in the database must be prioritized based on the main business objectives of the organization. For example, the human resources (HR) schema is usually the primary asset in most organizations. So any application must be linked to this schema in order to provide services to the corresponding employee.

- *Applications decoupling*: every application must provide only its basic functionality without mixing its concern with other application business. In addition, applications must not be aware of any other applications in the system. Their main task is to receive requests, process them, and provide results. So, any hardcoded links between applications must be eliminated.

- *Application architecture*: applications must be well structured in the sense that their composing components can be identified and the relationship between them is defined. The architecture of the application can then be utilized to identify the computational components from the data and control exchange components.

- *Separation of concerns*: the functional components of an application must be distinct in the sense that their business logics are not interleaved. For instance, credential check functionality must not be mixed with data retrieval or computation algorithm functionality. Every concern must be separated in a modular way (i.e. component) so it cannot be confused with other functionalities of an application.

- *Standardization of interfaces*: software applications must be wrapped in a way that complies with the standard interface used across the various systems within an organization. The interface usually defines the standard data exchanging model and control topology that is common to all systems.

- *Dynamic binding*: this attribute needs to be satisfied in enterprise systems where software applications can be used differently as per process design. In fact, this feature promotes a wider level of integration between different systems that conform to a standard interface.

- *Integration mechanism*: applications need to expose their standard interfaces in a layer within the overall environment where reaching them can be facilitated. This is usually referred to as a mediator platform where requests can be managed in terms of scheduling, routing, and finding of applications, among other things.

- *Authority matrix*: a system might be accessed by many users, and everyone has their own privileges to execute specific functionality. This is a mandatory attribute that any enterprise system must effectively handle and manage.

- *Data warehouse*: some organizations may have multiple databases for different types of application. This may increase the administration and maintenance overheads. Moreover, this may conflict with the strategy adopted by the organization that needs to integrate their scattered systems. A single unified data source must therefore be employed that wraps all the different databases and exposes a single interface to the applications. This approach is advantageous in the case of having various database types (e.g. Oracle, SQL, MySql).

To the best of our knowledge, these attributes are the most significant ones that organizations must consider when defining their system architecture. The identified attributes are the main driver for establishing our proposed architectural framework, which is given in the next section.

## 4. Proposed Enterprise System Architecture

One key driver for establishing our framework is the representation of workflow within a software system. Currently many systems develop their business processes hardcoded into the source code. So, whenever new business processes are required to be implemented the overall code must be modified. Moreover, applications are integrated in a one-to-one manner by writing *glue code* to establish the integration. This glue code is usually written as a mediator between two applications. Although this approach might look simple to some developers, it causes process design to become totally confused and mixed. In some cases glue code is injected into one of the applications themselves. This worst scenario as it will result in very tangled code that cannot be managed over the years.

Our proposed framework considers SOA [5] as an integration facilitator mechanism and not as a service delivery mechanism. The framework is composed of different layers that, we believe, any enterprise solution in the market must satisfy in order to ensure flexibility and extensibility of their systems. Figure 1 presents our proposed architecture for an enterprise solution.

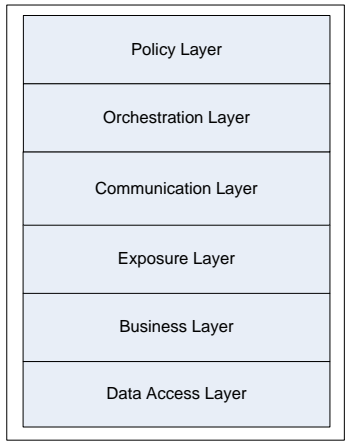

Fig. 1 Architectural Layers of Enterprise Solution

Each layer is independent of the other surrounding layers in terms of their main functionality. The description of these layers is as follows:

- *Data Access Layer*: this layer is responsible for managing the interaction between application and database and hiding the databases used in the organization. So, if different database technologies are used (e.g. SQL, Oracle), this layer will manage the connectivity with the corresponding source.

- *Business Layer*: this layer is responsible for executing the basic functionality that represents an organization's business needs. In the context of an ERP solution, this layer represents the fundamental modules offered by the solution such as HR, Finance, Projects, and Sales. Every one of these modules must be a standalone application that is not aware of any other modules.

- *Exposure Layer*: this layer is responsible for exposing the available applications from the application layers into services (e.g. web services, com components). All applications are therefore decoupled from their underlying environment and made available through request-response interaction mode.

- *Communication Layer*: the integration layer is responsible for establishing the communication pattern and routing protocols that enable service discovery and interaction. It defines the policies that comply with the standards adopted by vendors. For example, web services interact by exchanging SOAP messages over HTTP protocol. Any interaction between services must be accomplished through this layer. This is usually referred to as the Enterprise Service Bus (ESB) layer.

- *Orchestration Layer*: this layer defines the business processes that are employed by an organization. It is responsible for establishing the sequence by which services are going to be invoked to satisfy business requirements. For example, an attendance service might need to issue a request to a finance service to deduct a certain amount from an employee salary.

- *Policy Layer*: this layer is responsible for defining the privileges for accessing services. A different level of access rights can therefore be granted at this layer according to the defined policy.

The identified layers are not interchangeable as they must build up in a bottom-up manner. So, for example, a database can be established and tables created for an ERP system. Then, a number of standalone applications are developed on top of these tables to utilize the data in the tables. These applications must then be exposed in a standard manner in order to facilitate their integration with other applications to achieve new business needs. So the new exposed interfaces are pooled and made ready for requests. Workflows can then be defined on top of the available pool of services in order to integrate different applications seamlessly without affecting each application's concern. In fact, a workflow defines the design of a system where different components can be executed in a pre-defined sequence. Once all the business requirements are established (i.e. all functionality is implemented), there should be privileges assigned to personnel who are authorized to execute certain processes in the system.

## 5. ERP Solutions Analysis

A number of well-known ERP solutions are available nowadays in the market. Oracle, for instance, is among the prominent vendors in this field through their Oracle Apps, or the E-Business suite (EBS) [1]. Oracle ERP is a three-tier system that is composed of four basic modules, namely Human Resources, Project Management, Finance, and Asset Management. These modules are built on top of a unified Oracle database. The interaction between these modules is achieved via the Business Event System (BES) that triggers message creation or consumption of any registered parties. Oracle currently offers an additional package, namely the SOA suite, which can be integrated with the E-Business suite in order to promote enhanced scalability. ERP applications can therefore be exposed on the Oracle Service Bus (OSB) as services. These services then interact with each other through a business process design defined in BPEL. Recently, the key features of the SOA suite became an integral part of the Oracle E-business suite R12.1 package with the inclusion of the Oracle EBS adapter which exposes pl/sql as services. However, these added features are sold with different licenses which can be very expensive to some organizations, especially those in the government sector.

Microsoft offers a number of ERP solutions to suit various customer needs, one of which that is known as a comprehensive solution is Dynamics AX [2]. It employs the three-tier architectural pattern, namely, client tier, Application Object Server (AOS) tier, and database tier. The client contains forms and reports code. AOS is used to execute application objects such as classes and queries. The database is normally used to store data for the ERP. Microsoft Dynamics AX utilizes the Application Integration Framework (AIF) to facilitate the integration of application-to-application and also business-to-business. AIF supports the creation of generic web services and also document services; it also facilitates the consumption of external web services from within Dynamic AX. Another ERP solution provided by Microsoft is the Dynamics GP, which is also based on a three-tier architectural pattern. The application tier is composed of three main components: the Dexterity tool and runtime, Dynamics Application Dictionary, and SQL server. The Dexterity tool is used to build the forms and also to attach scripting code using sanScript to applications. The Dexterity runtime environment is used to enable the execution of a functioning application to end-users. This tool is therefore responsible for the development and the execution of the application interfaces. The Dynamics Application Dictionary (DAD) is responsible for storing the business logics in common component architecture such as COM+ and DCOM [6], so other distributed applications can use them as service providers. The main design consideration of this dictionary is to separate the presentation logic from the actual business logic of an application, so services can be accessed independently of any form or application of the presentation layer. The workflow engine is not part of the overall structure but Dynamics utilizes SharePoint to provide this feature.

SAP ERP [3], known as SAP R/3, is another prominent solution in the market. It is primarily based on a three-tier architectural style: the presentation layer, the application layer, and the database layer. The presentation layer represents a tiny application, namely sapgui.exe, that is usually installed on the client's machine. The application servers, namely SAP Netweaver, host different SAP services that execute code written in APAB/4 language. A messaging server is responsible for routing requests between applications and establishing a means of interaction between them. The main modules exhibited by SAP ERP are: Financials and Controlling (FICO), Human Resources (HR), Materials Management (MM), Sales & Distribution (SD), and Production Planning (PP).

It is apparent from the above that all the described ERP solutions provide similar kinds of functionality and also they share a common three-tier architectural pattern. The three-tier architectural pattern can satisfy, to some extent, the scalability requirement we described earlier; however, it is not very efficient in terms of integrating services or applications. Currently, the business logics are implemented in the application tier in all the ERP solutions. In Oracle ERP, some business logics are stored on a database as well. So when there is a need to integrate two or more applications or services together, there is a need to either modify part of the application's code or write an additional mediator application that establishes the linking between the corresponding parties. Therefore, our proposed solution to integrate workflow business in the context can tackle this problem and solidify the application layer. Moreover, it can satisfy the scalability and integration requirements identified earlier. The mapping of these solutions to our framework is given in Table 1 below.

Table 1: ERP solutions analysis

| Layer | Oracle ERP | Microsoft Dynamics | SAP R/3 |
|---|---|---|---|
| Data source | × | √ | √ |
| Business implementation | √ | √ | √ |
| Exposure | × | × | × |
| Communication | × | × | × |
| Policy | √ | √ | √ |
| Orchestration | × | × | × |

The Oracle ERP does not adopt the principle of data source where different types and technology of databases can be used, as it is restricted to its own technology platform. This is not the case in Microsoft Dynamics and also in SAP R/3 as the database link layer is developed to manage interconnectivity with any type of database servers. None of the ERP products adopt the notion of services where they decouple business logics from the underlying environment. Currently every application must be written in a specific programming language that sticks to certain architectural specification. This adds extra overheads when there is a future need for potential development. All the three ERP solutions lack a well-defined integration and communication layer that is responsible for managing interactions and also finding services. Microsoft Dynamics has a workflow engine that defines how documents must be flowing within an organization. However, the workflow engine is not designed to facilitate the orchestration and integration of applications or services.

It is obvious that all the ERP solutions focus mainly on the functional side to satisfy business needs; however, an architectural arrangement to support scalability and flexibility is not considered in the original building block of the system. These additional capabilities can be obtained for an enormous additional cost even though they play a significant role in enhancing the scalability and flexibility of software systems within organizations.

## 6. Case Study

We selected Umm Al-Qura University (UQU) as a case study for applying our framework as their environment is somewhat complicated to manage and control. We have worked at UQU in the IT deanship for more than three years. We observed, throughout this period, a number of challenges that hinder the university from fulfilling its mission. Some challenges are related to the functional capabilities of their systems while the majority relate to the processes and integration of different systems. Therefore, we decided to apply our study to the benefit of the university in order to comply with the new emerging business requirements.

Currently, one of the main objectives of UQU business is to establish a fully integrated environment that supports e-government business needs, so they need to have a rigorous solution that promotes changes without interrupting their daily working activities. Umm Al-Qura University established its information systems in early 1995 to serve around 3,600 employees and nearly 40,000 students at that time. It owns old-fashioned systems based on Oracle 6i for forms and reports that are built entirely on client-server pattern. The major functional systems include an in-house-built ERP, Student Information System (SIS), Library Information System (LIS), and Healthcare Information System (HIS). These systems are used today at the university to serve around 75,000 students and more than 7,000 employees with some enhancement to their functionality. However, software systems at UQU still lack many capabilities that become core-requirement nowadays in terms of compatibility with different environments (e.g. mobile devices) and also the services provided to students and faculty members in the University. Moreover, with the pioneering e-government movements within the region of Saudi Arabia, it becomes necessary that organizations apply major changes to their systems in order to accommodate these new requirements, one of which is process automation which solely requires splitting functional aspects of an application from the process aspects. Currently, modifications to add features to any of the systems are done in an ad-hoc manner where the application's code is modified to satisfy new business requirements. Specifically, business processes are implemented directly into the forms, confusing the functional aspects of an application with the non-functional ones. As a result, the complexity of UQU systems builds up rapidly in a manner that will become very hard to manage in the near future.

Our analysis of the main technologies used at UQU revealed that it currently has three different environments: SharePoint, PHP, and Oracle. Our proposed architecture is meant to integrate all systems regardless in a technology-neutral manner. The proposed system architecture for UQU is given in Figure 2 below.

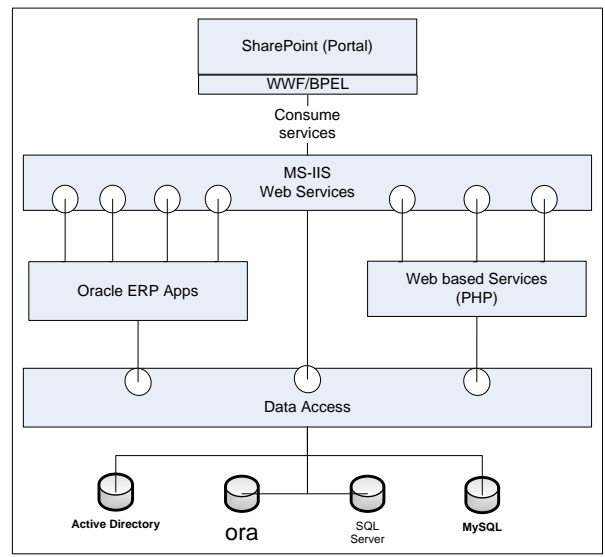

Fig. 2 UQU's Proposed System Architecture

The figure illustrates the proposed architecture for satisfying the business need of UQU based on the resources that are currently available to the university. The main objective of this solution is to promote a fully integrated environment that facilitates internal and external data exchange, in addition to promoting scalability for future development. UQU currently owns a full package of SharePoint 2010, an in-house built Oracle ERP solution, a website and a number of services in PHP, and an Internet

Information Server (IIS). In our proposed solution, SharePoint is utilized to play two main roles: the web presence and the service orchestration layer where business processes are defined through windows workflow foundations (WWF) provided by the SharePoint workflow engine. Services are exposed to SharePoint through the Microsoft-IIS layer where web services are defined. Therefore every application must be wrapped and exposed as a standalone web service that can be consumed by SharePoint. This capability simulates the basic functionality of an ESB for service integration and management which represent the communication layer for integrating the various applications in an organization. SharePoint 2010 must work only on an SQL server, hence, in this solution, we propose using the SQL server for document flow management purposes without interfering with the university database by any means. The resulting architecture should promote a high degree of extensibility and flexibility where different business processes within or between departments become composable and fully automated.

## 7. UQU Systems Migration Guideline

We referred partially to the SMART process [18] to help us examining the feasibility of migrating UQU legacy systems into the new SOA based environment. The analysis uncovers a number of activities that need to be conducted in order to implement the proposed solution, they are:

- Re-factor applications in order to eliminate potential decoupling applications from each other so everyone can provide its standard set of functionality without any reference to other applications in the system.

- Extract stored PL/SQL procedures in the Oracle DB and wrap them with containers to be exposed as web services.

- Business logic must be separated from the Oracle forms by following the Model-View-Controller (MVC) architectural pattern. So, business logics can be accessed from different views and not restricted to a single usage. This might be achieved through the migration to the ADF. So, extract the source code from oracle forms and encapsulate them in a well-defined business component (BC) models that can be invoked directly by forms. Thus, functionality that is embedded in forms can be de-coupled in self-contained components.

- Establish the linkage between forms, BC web-service, and PL/SQL web-service. So, forms can be hardcoded to invoke BC services. However, BC services must interact with the PL/SQL services via a defined work flow in order to support dynamic binding. So, no code must be used to establish the linkage between services.

- The resulted web services must be exposed through Microsoft-IIS that establishes messages routing protocol between web services. The Microsoft-IIS is considered as the service layer in this scenario.

- Active Directory must be integrated to the service layer in order to provide credential check and assign basic privileges to users according to their pre-defined profiles.

- Utilize the workflow (WF) engine provided by SharePoint 2010 in order to implement business processes. The implemented WF represents the main thread of control that establishes the design for consuming the exposed services. So, services can be placed and executed in a sequence to fulfill business requirements.

- The functional interface must be separated from the architectural interface [19]. So, UQU team must identifying the business logic such as data link, connectors, and modules life-cycle control code and separate them from the core functional business logic. This helps to identify the potential functional services that can be consumed directly by clients and separate them from any supporting services that may be related to the architecture of the legacy system.

The above set of activities describes how UQU can migrate their current applications to satisfy SOA basic requirements. These activities are considered with the assumption that UQU is going to utilize the current Oracle application not only as black boxes but as components that are not going to be modified in further.

## 8. Conclusion

This paper presented our proposed framework to evaluate enterprise solutions in the market. The framework is based primarily on the concept of SOA to define the different architectural layers. Although this study was limited only to three ERP solutions in the market, these solutions are the most commonly known ones in the Saudi Market. The paper has drawn organizations' attention to the idea of investing in the process of defining software architecture for their systems in order to generate a reference model to fit different technology in the market to their business needs. The next step in this research is to implement the potential migrating roadmap resulted from this work to migrate the current systems at UQU to comply with the defined framework.

## Acknowledgments

The author would like to thank the IT dean Dr. Fahad Al-Zahrani for supporting and encouraging this work. Also special thanks to Eng. Ashraf Asfour (the head of Application Development Department at UQU) who helped in preparing data used in this research and facilitating accessibility to the underlying system architecture. This work would not be possible without the support of Umm Al-Qura University.

**Basem Y. Alkazemi** is an assistant professor at Umm Al-Qura University (UQU) in Saudi Arabia under the school of computer science & Engineering. He obtained his PhD in 2009 from Newcastle University in U.K. His PhD topic was concerned with addressing the complexity of re-using open-source software components. Basem is currently holding the position of vice dean of IT deanship for e-government at UQU. One of his main duties is to establish a framework that leads to the integration of all the university software systems in a unified model. He is a member in the IEEE, SIGSOFT-ACM, and SEI societies. His main research interests include software architectural patterns, software product lines, Aspect-oriented SE, SOA, and CBSE.